\documentclass[conference,10pt,letterpaper]{IEEEtran}
\IEEEoverridecommandlockouts
\usepackage{cite}
\usepackage{amsmath,amssymb,amsfonts, mathdots, bm}
\usepackage{algorithmic}
\usepackage{graphicx}
\usepackage{textcomp}
\usepackage{xcolor}
\usepackage{color}
\usepackage{siunitx}

\usepackage{pgf, tikz, pgfplots}
\pgfplotsset{compat=1.18}
\usepackage{circuitikz}
\usetikzlibrary{shapes,arrows, calc, fit}
\usetikzlibrary{positioning}
\def\BibTeX{{\rm B\kern-.05em{\sc i\kern-.025em b}\kern-.08em
    T\kern-.1667em\lower.7ex\hbox{E}\kern-.125emX}}
\usepackage[numbers]{natbib}
\usepackage{adjustbox}
\usepackage{svg}
\usepackage{lipsum}
\usepackage{natbib}

\definecolor{kit-blue100}{cmyk}{.8,.5.,0,0}
\definecolor{kit-blue70}{cmyk}{.56,.35,0,0}
\definecolor{kit-blue50}{cmyk}{.4,.25,0,0}
\definecolor{kit-blue30}{cmyk}{.24,.15,0,0}
\definecolor{kit-blue15}{cmyk}{.12,.075,0,0}

\definecolor{kit-green100}{cmyk}{1,0,.6,0}
\definecolor{kit-green70}{cmyk}{.7,0,.42,0}
\definecolor{kit-green50}{cmyk}{.5,0,.3,0}
\definecolor{kit-green30}{cmyk}{.3,0,.18,0}
\definecolor{kit-green15}{cmyk}{.15,0,.09,0}

\definecolor{KITgreen}{rgb}{0,.59,.51}

\definecolor{KITpalegreen}{RGB}{130,190,60}
\colorlet{kit-maigreen100}{KITpalegreen}
\colorlet{kit-maigreen70}{KITpalegreen!70}
\colorlet{kit-maigreen50}{KITpalegreen!50}
\colorlet{kit-maigreen30}{KITpalegreen!30}
\colorlet{kit-maigreen15}{KITpalegreen!15}

\definecolor{KITblue}{rgb}{.27,.39,.66}

\definecolor{KITyellow}{rgb}{.98,.89,0}
\definecolor{kit-yellow100}{cmyk}{0,.05,1,0}
\definecolor{kit-yellow70}{cmyk}{0,.035,.7,0}
\definecolor{kit-yellow50}{cmyk}{0,.025,.5,0}
\definecolor{kit-yellow30}{cmyk}{0,.015,.3,0}
\definecolor{kit-yellow15}{cmyk}{0,.0075,.15,0}

\definecolor{KITorange}{rgb}{.87,.60,.10}
\definecolor{kit-orange100}{cmyk}{0,.45,1,0}
\definecolor{kit-orange70}{cmyk}{0,.315,.7,0}
\definecolor{kit-orange50}{cmyk}{0,.225,.5,0}
\definecolor{kit-orange30}{cmyk}{0,.135,.3,0}
\definecolor{kit-orange15}{cmyk}{0,.0675,.15,0}

\definecolor{KITred}{rgb}{.63,.13,.13}
\definecolor{kit-red100}{cmyk}{.25,1,1,0}
\definecolor{kit-red70}{cmyk}{.175,.7,.7,0}
\definecolor{kit-red50}{cmyk}{.125,.5,.5,0}
\definecolor{kit-red30}{cmyk}{.075,.3,.3,0}
\definecolor{kit-red15}{cmyk}{.0375,.15,.15,0}

\definecolor{KITpurple}{RGB}{160,0,120}
\colorlet{kit-purple100}{KITpurple}
\colorlet{kit-purple70}{KITpurple!70}
\colorlet{kit-purple50}{KITpurple!50}
\colorlet{kit-purple30}{KITpurple!30}
\colorlet{kit-purple15}{KITpurple!15}

\definecolor{KITcyanblue}{RGB}{80,170,230}
\colorlet{kit-cyanblue100}{KITcyanblue}
\colorlet{kit-cyanblue70}{KITcyanblue!70}
\colorlet{kit-cyanblue50}{KITcyanblue!50}
\colorlet{kit-cyanblue30}{KITcyanblue!30}
\colorlet{kit-cyanblue15}{KITcyanblue!15}

\begin{document}

\title{Fractional Chirp-Slope-Shift-Keying for SDR-based Search and Rescue Applications
}

\author{Daniel Gil Gaviria, Marcus Müller, Felix Artmann and Laurent Schmalen\\
\IEEEauthorblockA{Communications Engineering Lab (CEL), Karlsruhe Institute of Technology (KIT) \\ Hertzstr. 16, 76187 Karlsruhe, Germany, Email: \texttt{daniel.gil@kit.edu}}
}

\maketitle

\begin{abstract}
The use of modern software-defined radio (SDR) devices enables the implementation of efficient communication systems in numerous scenarios. Such technology comes especially handy in the context of search and rescue (SAR) systems, enabling the incorporation of additional communication data transmission into the otherwise sub-optimally used SAR bands at 121.5 and 243~MHz. In this work, we propose a novel low-complexity, energy-efficient modulation scheme that allows transmission of additional data within chirped homing signals, while still meeting the standards of international SAR systems such as COSPAS-SARSAT. The proposed method modulates information onto small deviations of the chirp slope with respect to the required unmodulated chirp, which can be easily detected at the receiver side using digital signal processing. 
\end{abstract}

\section{Introduction}
COSPAS-SARSAT is an international satellite-aided search and rescue system for the detection and localization of distress beacons, which yearly aids several hundreds of rescue missions facilitating the saving of thousands of lives. The sequence of events of the COSPAS-SARSAT system is roughly structured as follows \cite{Zurabov1998,Lev1993}: after the occurrence of a distress, an emergency locator transmitter (ELT) carried by vehicles or individuals sends a distress signal to the COSPAS-SARSAT satellite constellation. The signal is retransmitted in real time to ground stations which alert local mission control centers and trigger a search and rescue (SAR) mission. Additionally, the ELT emits a homing signal carrying no communication data, which serves as a reference to be located by the rescue mission \cite{R-REC}. The specifications for this signal are described in detail below. 

Apart from the distress signal, no further transmission of data is foreseen in the original standard. However, real-time communications about the details of the distress might be of great relevance for a successful SAR mission. Descriptions of the location of the distress, the number of persons involved and further situation-specific details might ease and accelerate the mission. Some information can be transmitted through the satellite up-link. For instance, a software-defined radio (SDR) based data transmission through the initial satellite uplink using direct sequence spread spectrum (DSSS) was introduced in \cite{Mla2022}. Nevertheless, the low rate and high latency of the satellite link might be prohibitive for time-critical rescue maneuvers in distress scenarios. The possibility of an additional communication link within the homing signal still offers the potential for the transmission of valuable information.

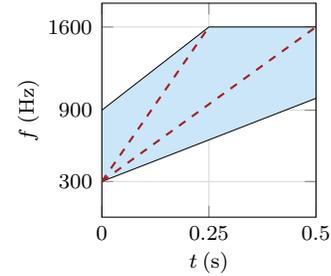
\begin{figure}
  \centering
\begin{tikzpicture}[baseline]

    \begin{axis}[
        width=0.5\columnwidth,
        height=0.5\columnwidth,
          grid=major, %
          grid style={gray!30}, %
          axis background/.style={fill=white},
          ylabel=$f\: (\mathrm{Hz})$,
          y label style={at={(-.25,0.5)},font=\footnotesize},
          xlabel=$t\: (\mathrm{s})$,
          x label style={at={(.5,-0.12)},font=\footnotesize},
          enlarge x limits=false,
          enlarge y limits=false,
          xmin = 0,
          xmax=0.5,
          ymax = 1800,
          ymin = -0,
          xtick={0,0.25,0.5},
          xticklabels={$0$, $0.25$, $0.5$},
          ytick={300,900,1600},
          yticklabels={$300$,$900$, $1600$},
          legend style={at={(1,1)}, font=\normalsize, fill opacity=0, text opacity=1, column sep=0.2cm},
          legend cell align={left},
          tick label style={font=\footnotesize}
        ]
        \addplot[fill=kit-cyanblue30,line width= 0pt] table {
        0 300
        0 900
        0.25 1600
        0.5 1600
        0.5 1000
        }--cycle;
       \addplot[KITred, dashed, thick] coordinates {
      (0,300)(0.25,1600)
       };
       \addplot[KITred, dashed, thick] coordinates {
        (0,300)(0.5,1600)
         };
      
    \end{axis}
    \end{tikzpicture}

    \caption{Available time and frequency range for a chirp repetition according to standard specifications \cite{R-REC} (blue) and exemplary chirp realizations (red-dashed)} 
    \label{fig:specs_sketch} 
\end{figure}

The COSPAS-SARSAT standard specifications aim to shape the homing signal as an AM-modulated audio signal with a characteristic chirped sound that can be rapidly localized by the rescue mission. Its main characteristics are given in \cite{R-REC} and can be summarized as follows:

\begin{itemize}

\item The emission should include the carrier frequency amounting to at least \SI{30}{\percent} of the power at all times.
\item The emission consists of an audio frequency sweep within a bandwidth $B_0$ larger than~\SI{700}{ \hertz} with a minimum and maximum frequency being in the range \SI{300}{\hertz} to \SI{1600}{\hertz} and a sweep rate of  $R\in\left[2,4\right]$ repetitions per second. 

\end{itemize}
The resulting available range in time and frequency according to these constraints and two possible realizations of the homing signal are sketched in Fig. \ref{fig:specs_sketch}.

The specifications also explicitly leave room for any type of modulation as long as it does not impair the precise localization of the radio beacon. Several approaches to chirp spread spectrum (CSS) have been proposed in the past,  especially in the context of long-range communication \cite{Baruffa2022}. The purpose of CSS modulation is usually to spread fast symbols over the time and bandwidth of a chirp. The specified limited bandwidth and a low chirp repetition render CSS unsuitable for the application highlighted in this paper since only one symbol is represented by the constant slope, phase, relative frequency or time shift of each chirp. Only low bit rates can thus be achieved using these methods within the given signal specifications.   To overcome this limitation, we propose fractional chirp slope shift keying, which consists of modulating information data onto the instantaneous slope and frequency of fractions of the chirp allowing higher data rates for efficient communication.

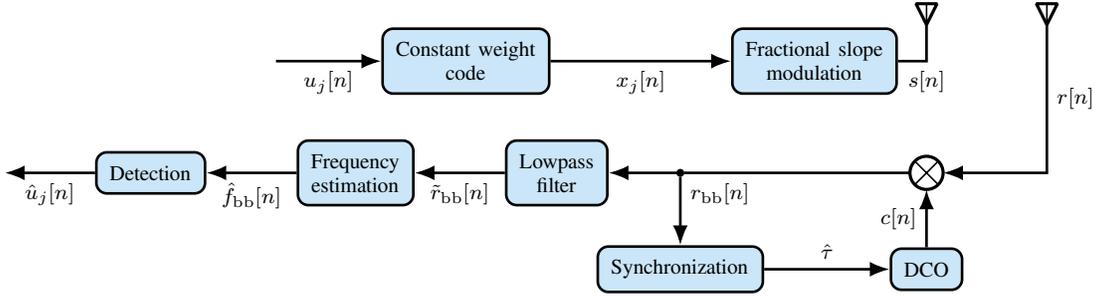
\begin{figure*}[h!]
    \centering
    \centering
\tikzset{
    ultra thin/.style= {line width=0.1pt},
    very thin/.style=  {line width=0.2pt},
    thin/.style=       {line width=0.4pt},%
    semithick/.style=  {line width=0.6pt},
    thick/.style=      {line width=0.8pt},
    very thick/.style= {line width=1.2pt},
    ultra thick/.style={line width=1.6pt}
}

    \tikzstyle{bl} = [draw, line width=1pt, rectangle, inner sep=.5em, rounded corners, fill=kit-cyanblue30, font=\footnotesize]
    \tikzstyle{bl2} = [draw, very thick, circle, inner sep=.8em]
     \tikzstyle{input} = [coordinate]
     \tikzstyle{output} = [coordinate]
     \tikzstyle{every node}=[line width=1pt, font=\footnotesize]
     \tikzstyle{dot}=[circle,fill,draw,inner sep=0pt,minimum size=2pt]
    \tikzstyle{sum}=[draw, circle, minimum size=10pt]
     
    \begin{circuitikz}[auto, line width=1pt, x=0.2cm,y=0.35cm]

    \node[input, name=input]{};
    \node[bl, name=manCode, align=center, right =7 of input]{Constant weight\\ code};
    \node[bl, name=mod, align=center, right = 12 of manCode]{Fractional slope \\ modulation};
    \node[bareantenna,line width=0.6pt, scale=0.4, above right= 0.1 and 1.2 of mod, name=ant_out] {};
    \node[mixer, line width=0.6pt, scale=0.45, below=5 of ant_out,name=mix]{};
    \node[dot, left=15 of mix, ](synchdot){};
    \node[bl, left=4.5 of synchdot, align=center, name=lp ]{Lowpass \\ filter};
    \node[bl, left= 6 of lp,name=pll, align=center]{Frequency \\ estimation};
    \node[bl, left=6 of pll, name=mf, align=center]{Detection};
    \node[output, left=6 of mf, name=output]{};
    
    \node[bl, below= 2.6 of synchdot](synch){Synchronization};
    \draw (synch-|mix) node[bl,  align=center](dco){DCO};

     \draw [arrows = {-Latex[width = 2mm]} ] (input) -- node[name=u, below] {$u_j[n]$} (manCode);
     \draw [arrows = {-Latex[width = 2mm]} ] (manCode) --node[name=x, below] {$x_j[n]$} (mod);
     \draw [-] (mod) -- (mod-|ant_out)node[name=s, below]{$s[n]$} -- (ant_out);
     \draw [-] (mix) -- (synchdot)node[below right]{$r_\mathrm{bb}[n]$};
     \draw [arrows = {Latex[width = 2mm]-} ](mix)-- ++(8,0) coordinate(temp) -- node[right]{$r[n]$} (temp|-ant_out)coordinate(temp);
     \draw (temp) node[bareantenna, line width=0.6pt, scale=0.4, anchor=bottom]{};
     \draw [arrows = {-Latex[width = 2mm]} ] (synchdot) -- (lp);
     \draw [arrows = {-Latex[width = 2mm]} ] (lp) --  node[below]{$\tilde{r}_\mathrm{bb}[n]$}(pll);
     \draw [arrows = {-Latex[width = 2mm]} ] (pll) -- node[name=rbb, below] {$\hat{f}_\mathrm{bb}[n]$} (mf);
     \draw [arrows = {-Latex[width = 2mm]} ] (mf) -- node[name=rbb, below] {$\hat{u}_j[n]$} (output);

     \draw[arrows = {-Latex[width = 2mm]} ] (synchdot) -- (synch);
     \draw[arrows = {-Latex[width = 2mm]} ] (synch) -- node[above]{$\hat{\tau}$} (dco);
     \draw[arrows = {-Latex[width = 2mm]} ] (dco)-- node[left]{$c[n]$} (mix);
        
    \end{circuitikz}
    \caption{Block diagram of the complete transmitter and receiver systems}
    \label{fig:block_diagram}
\end{figure*}

\begin{center}
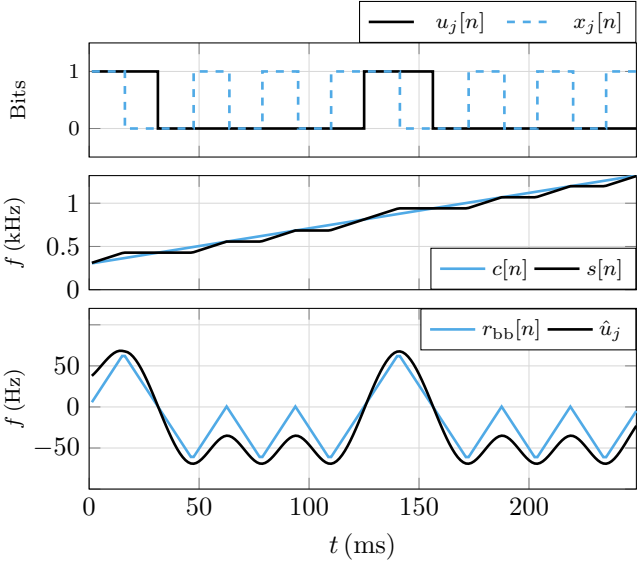
\begin{figure}[!tb]
  \hspace*{-.25cm}  
\begin{tabular} {rl}
    \begin{tikzpicture}[baseline]
   
    \begin{axis}[
        width=1\columnwidth,
        height=0.35\columnwidth,
        axis background/.style={fill=white},
          grid=major, %
          grid style={gray!30}, %
          ylabel=$\mathrm{Bits}$,
          y label style={at={(-0.1,0.5)}},
          enlarge x limits=false,
          enlarge y limits=false,
          xmin = 0,
          ymax = 1.5,
          ymin = -0.5,
          xtick={0,.050,.100,.150,.200,.250},
          xticklabels={},
          ytick={0,1},
          yticklabels={$0$,$1$},
          legend columns=-1,
          legend style={at={(1,1)}, anchor=south east,  fill opacity=1, text opacity=1, column sep=0.2cm,font=\footnotesize},
          legend cell align={left},
          tick label style={font=\footnotesize},
          y label style={font=\footnotesize}
        ]
        \addplot+[mark=none, color=black, line width= 1pt]
        table[x=t,y=b,col sep=comma] {signals.csv};
        \addlegendentry{$u_j[n]$ } 
        \addplot+[mark=none, color=KITcyanblue, line width= 1pt, dashed]
        table[x=oldt,y=man,col sep=comma] {signals.csv};
        \addlegendentry{$x_j[n]$ }
    \end{axis}
    \end{tikzpicture}
\\
    \begin{tikzpicture}[baseline]
    \begin{axis}[
        width=1\columnwidth,
        height=0.35\columnwidth,
        axis background/.style={fill=white},
          grid=major, %
          grid style={gray!30}, %
          ylabel= $f \,(\mathrm{kHz})$,
          y label style={at={(-0.1,0.5)}},
          xtick={0,.050,.100,.150,.200,.250},
          xticklabels={},
          ytick={0, 500,1000,1500},
          yticklabels={$0$, $0.5$, $1$,$1.5$},
          enlarge x limits=false,
          enlarge y limits=false,
          xmin = 0,
          ymin = 0.00,
          legend columns=-1,
          legend style={at={(1,0)},anchor=south east, fill opacity=1, text opacity=1,font=\footnotesize},
          legend cell align={left},
          y label style={font=\footnotesize}
        ]
        \addplot+[mark=none, color=KITcyanblue, line width= 1pt]
        table[x=t,y=f2,col sep=comma] {signals.csv};
        \addlegendentry{$c [n]$}
       \addplot+[mark=none, color=black, line width= 1pt]
        table[x=t,y=f,col sep=comma] {signals.csv};
        \addlegendentry{$s [n]$}
        
    \end{axis}
    \end{tikzpicture}
    \\
    \begin{tikzpicture}[]
    \begin{axis}[
        width=\columnwidth,
        height=0.45\columnwidth,
        axis background/.style={fill=white},
          grid=major, %
          grid style={gray!30}, %
          xlabel= $t \, (\mathrm{ms})$ ,
          ylabel= $f \,(\mathrm{Hz})$,
          y label style={at={(-0.1,0.5)}},
          enlarge x limits=false,
          xmin = 0,
          xtick={0,.050,.100,.150,.200,.250},
          xticklabels={$0$,$50$,$100$,$150$,$200$,$250$},
          ytick={-100,-50,0,50,100},
          yticklabels={ ,$-50$,$0$,$50$,},
          ymax = 120,
          ymin = -100,
          legend columns=-1,
          legend style={at={(1,1)},anchor=north east,  fill opacity=1, text opacity=1,font=\footnotesize},
          legend cell align={left},
          y label style={font=\footnotesize}
        ]
        \addplot+[mark=none, color=KITcyanblue, line width= 1pt]
        table[x=t,y=deltaf,col sep=comma] {signals.csv};
        \addlegendentry{$r_\mathrm{bb}[n]$}
        \addplot+[mark=none, color=black, line width= 1pt]
        table[x=t,y=corr,col sep=comma] {signals.csv};
        \addlegendentry{$\hat{u}_j$}
    \end{axis}
    \end{tikzpicture}
    
\end{tabular}
    \caption{Time domain signals for $B_0=\SI{1}{\kilo\hertz}$, $R=\SI{4}{\hertz}$ and \SI{32}{\bit/\second} for an exemplary bit stream $u_j$}
    \label{fig:signals}
\end{figure} 
\end{center}
\vspace{-1cm}

In the remainder of this paper, the modulation and transceiver system will be explained in detail in Sec.~\ref{sec:system}. In Sec. \ref{sec:blocks}, the individual building blocks will be studied and the achievable performance will be derived in Sec. \ref{sec:CRB}. Finally, the results are evaluated for realistic scenarios through simulations in Sec. \ref{sec:results}.

\section{System Concept} \label{sec:system}
In its original conception, the homing signal corresponds to a non-information carrying time-discrete linear chirp

\begin{align} \label{eq  :constant_slope}
    &c\!\left[ n\right]=\mathrm{exp}\!\left( \frac{\mathrm{j}\pi k_0n^2}{2} \right), & 0\leq n < N\mathrm{,}
\end{align}
where $N$ is the number of samples per chirp at a set sampling frequency $f_\mathrm{s}$, $k_0$ denotes the constant slope of the chirp and $\mathrm{j}=\sqrt{-1}$. The bandwidth conditions are met if $k_0=B_0/N$ holds.

In order to add information onto this signal, we propose a modulation scheme in which symbols are keyed as a variable instantaneous slope $\kappa[n]$ for the duration of a symbol, which can be shorter than the repetition period of a chirp.

Let $x_j[n]\in \{0,1\} $ for $jM \leq n < (j+1)M$ be the binary sequence to be modulated, where $M$ is the sample length of an information bit for a given bit rate. The recursively defined instantaneous phase $\varphi[n]$, instantaneous frequency (IF) $f[n]$ and the resulting modulated signal $s[n]$ are then generated by 
\begin{align}\label{eq:definitions}
  s[n]&=\mathrm{exp}\left( \mathrm{j}\varphi[n] \right) \text{,}  \\
  \varphi[n]&= \varphi[n-1]+2\pi f[n]T_s  \text{,}\\
  f[n] &= f[n-1]+ \kappa_{x_j[n]} \text{,}
\end{align}
where $\kappa_0 \neq \kappa_1$ are the instantaneous chirp slopes corresponding to each binary symbol $x_j[n]$.

The total bandwidth of an information-carrying chirp then amounts to the sum of the instantaneous slopes over its duration. Thus, in order to keep the specified total bandwidth~$B_0$ of a complete chirp constant,  the condition 
\begin{equation}\label{eq:bw_condition}
    \sum_{n=0}^{N-1} \kappa_i[n]=B_0
\end{equation}
must hold. A constant-weight code can ensure that this condition is met independently of the unknown information bits.

In order to recover the information bits at the receiver, the difference between the information carrying received chirp $r[n]$ and a synchronized linearly modulated chirp $c[n]$  is determined. This can be achieved by downconversion through mixing both signals, lowpass filtering of the mixing products and a subsequent IF estimation of the resulting baseband signal. 

The block diagram of the complete transmitter and receiver concepts and the most relevant signals in the time domain are shown in Fig. \ref{fig:block_diagram} and Fig. \ref{fig:signals} respectively. Each step of the process will be explained in the following section and the performance of the complete system will be evaluated afterwards.

\section{Building Blocks}\label{sec:blocks}
\subsection{Constant-weight Codes}
 A constant-weight code is a mapping  $c:\mathbb{F}_2^p \rightarrow \mathbb{F}_2^q$ of a binary information sequence of length $p$ into a codeword of length $q$ with a constant Hamming weight $w$. The constant-weight code is defined as:

\begin{equation}
    \mathcal{C}=\left\{\bm{x} \in \mathbb{F}_2^q : d_\mathrm{H}(\bm{x})=w   \right\}\text{,}
\end{equation}
where $d_\mathrm{H}$ is the Hamming weight of $\bm{x}$.

In this paper, we consider two different constant-weight codes: the Manchester and 6b8b codes. First, we study the Manchester code, for which ${p=1}$, ${q=2}$ and ${w=1}$ holds. Let $\mathbf{u}\in \mathbb{F}_2^p$ be an arbitrary information bit stream which is oversampled by a factor of $M$. For each oversampled information bit $u_j[n]\in\{0,1\}$, the sequence $x_j[n]$ is given by
\begin{align}\label{manchester}
    x_j[n]=   
\begin{cases}
	 u_j\text{,}    & \text{for }  0\leq n < M/2,  \\
  \bar{u}_j\text{,}  & \text{for }  M/2 \leq n < M,
\end{cases}    
\end{align}
where $\bar{u}_j$ is the negated bit value of $u_j$.  Note that the duration of the coded bits is set to $M/2$ samples in order to keep the information bit rate unmodified. In other words, each information bit $u_j$ is flipped after half of its duration. Furthermore, $\kappa_0$ and $\kappa_1$ are set to $0$ and $2k_0$ respectively, in order to fulfill  (\ref{eq:bw_condition}).

Despite having a code rate of $1/2$, the Manchester code is a well-suited option for this specific application for two main reasons: first, it leads to the minimum possible deviation from the specified linearly modulated chirp and second, information bits can be easily detected at the receiver by evaluating the difference between the IF $f_\mathrm{bb}[n]$ of the received signal $r[n]$ and the local constant slope chirp signal $c\!\left[ n \right]$, as will be shown below.

The IF of $\tilde{r}_\mathrm{bb}$ for the duration of an information bit can be expressed as
\begin{equation}\label{eq:triangles}
    f_\mathrm{bb}[n] =\pm
	\begin{cases}
		k_0 n, &  \text{for } jM\leq n < (j+\frac{1}{2})M,\\
		k_0 M - k_0 n, & \text{for }  (j+\frac{1}{2})M \leq n < (i+1) M \mathrm{ .}
	\end{cases}       
\end{equation}

 Since the triangular form of the IF of $\tilde{r}_\mathrm{bb}[n]$ is expected to follow the shape in (\ref{eq:triangles}) for every information bit, detection can be performed through a matched filter given by ${g[n]=\left|{f_\mathrm{bb}[n]}\right|}$. %

Additionally, we implement a 6b8b code which maps binary sequences of length $p=6$ onto $2^6$ codewords $\bm{x}$ of length $q=8$ with constant Hamming weight $w=4$ \cite{widmer2005dc}. 

For the fractional slope modulation, the duration of the coded bits is set to $(3M/4)$ to keep the information bit rate unmodified and the instantaneous slopes $\kappa_0$ and $\kappa_1$ corresponding to each coded bit are again set to $0$ and $2k_0$ in order to fulfill the bandwidth condition (\ref{eq:bw_condition}).

In contrast to the Manchester code, the 6b8b code also allows series of up to four consecutive binary 1s or 0s within coded sequence. This leads to a larger possible IF deviation from a linearly sloped chirp but still assures that the desired frequency range is covered within each codeword. 
In this case, decoding can be performed using a matched filter bank which determines the sequence $\bm{x}$ with the frequency modulated signal $f_{\bm{x}}$ that yields the maximal correlation with the received sequence. For an AWGN channel, this corresponds to a minimum squared error detector and can be expressed as \cite{proakis2008}:

\begin{equation}
    \hat{\bm{x}}=\arg \max_{\bm{x}\in \mathcal{C}} \mathrm{corr}(\hat{f}_\mathrm{bb},f_{\bm{x}})\mathrm{.}
\end{equation}

\subsection{Instantaneous Frequency Estimation}\label{Receiver}

An efficient and reliable estimation of the IF of $r_\text{bb}[n]$ is of great relevance for the performance of the presented system. Classical spectrogram approaches like short-time Fourier transform perform poorly in scenarios where the analyzed signal is not stationary within the duration of an observation window. Additionally, the resolution required to detect small frequency deviations on the baseband signal would demand prohibitively long observation windows. In this paper, we study two methods that overcome these limitations:

\subsubsection{Digital phase-locked-loop (DPLL) frequency estimation}

On one hand, the use of DPLL suits the purpose of estimating the IF of signals with time-varying frequency while requiring low computational effort. It corresponds to a recursive mean least squares approach (RMLS) \cite{Boashash}.

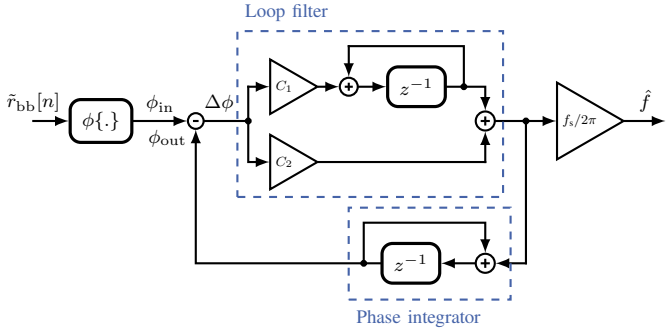
\begin{figure}[tb]
\centering
\tikzset{
    ultra thin/.style= {line width=0.1pt},
    very thin/.style=  {line width=0.2pt},
    thin/.style=       {line width=0.4pt},%
    semithick/.style=  {line width=0.6pt},
    thick/.style=      {line width=0.8pt},
    very thick/.style= {line width=1.2pt},
    ultra thick/.style={line width=1.6pt},
    gain/.style     = {draw, thick, isosceles triangle, minimum height = 0.8em,
        isosceles triangle apex angle=60}    
}
    \tikzstyle{dot}=[circle,fill,draw,inner sep=0pt,minimum size=2pt]
    \tikzstyle{sum}=[draw, circle, minimum size=10pt]
    \tikzstyle{bl} = [draw, very thick, rectangle, inner sep=.5em, rounded corners]
    \tikzstyle{bl2} = [draw, very thick, circle, inner sep=.2em]
    \tikzset{jump/.style args={(#1) to (#2) over (#3) by #4}{
        insert path={
            let \p1=($(#1)-(#3)$), \n1={veclen(\x1,\y1)},
            \n2={atan2(\y1,\x1)}, \n3={abs(#4)}, \n4={#4>0 ?180:-180}  in
            (#1) -- ($(#1)!\n1-\n3!(#3)$)
            arc (\n2:\n2+\n4:\n3) -- (#2)
        }
     }}
     \tikzstyle{input} = [coordinate]
     \tikzstyle{output} = [coordinate]
     \tikzstyle{every node}=[font=\footnotesize]
     \definecolor{KITblue}{rgb}{.27,.39,.66}
 \begin{tikzpicture}[auto, thick, scale=0.9, transform shape, node distance=1cm,line cap=rect]
    \node[input, name=input]{};
    \node[bl, name=phase, align=center, right of= input, node distance=1cm]{$\phi\{ .\}$};
    \node[circle, draw, name=sub, align=center, right of= phase, node distance=1.4cm, scale=0.7]{};
    \draw[-, thick] (sub.west)+(1mm,0) --  ($(sub.east)+(-1mm,0)$);
    \node[dot, draw, right= 0.6cm of sub](dot1){};
    \node [gain,scale=0.7, above right= 0.2cm and 1cm of sub](c1_mult){$C_1$};
    
    \node[circle, draw, name=sumz1, right=0.3cm of c1_mult, scale=0.8]{};
    \draw[-] (sumz1.west)+(1mm,0) --  ($(sumz1.east)+(-1mm,0)$);
        \draw[-] (sumz1.south)+(0,1mm) --  ($(sumz1.north)+(0,-1mm)$);
    \node[bl, align=center, right of= sumz1, node distance=1cm](delay1){$z^{-1}$};
    \node[dot, right of= delay1, node distance=0.7cm](dot2){};
    
    \node[gain,draw, name=c2_mult, scale=0.7, below right= 0.3cm and 1cm of sub] {$C_2$};

    \node[sum, right=3.3cm of dot1, scale=0.8](sum1){};

    \draw[-] (sum1.west)+(1mm,0) --  ($(sum1.east)+(-1mm,0)$);
        \draw[-] (sum1.south)+(0,1mm) --  ($(sum1.north)+(0,-1mm)$);
    \node[sum, below=1.8cm of sum1, scale=0.8](sum2){};
    \draw[-] (sum2.west)+(1mm,0) --  ($(sum2.east)+(-1mm,0)$);
    \draw[-] (sum2.south)+(0,1mm) --  ($(sum2.north)+(0,-1mm)$);
    \node[bl, left= 0.5cm of sum2](delay2) {$z^{-1}$};
    \node[dot, left=0.2cm of delay2](dot5){};

    \node[dot, right=0.4 cm of sum1] (dot4){};
    \node[gain,draw, right=0.4cm of dot4, scale=0.7] (mult_out){$f_\text{s}/2\pi$};
    \node[output,  name=output, right=0.6cm of mult_out]{};

  \draw[-latex] (input) node[name=x, left, above]{$\tilde{r}_\mathrm{bb}[n]$} --(phase);
  \draw[-latex] (phase) -- node[above]{$\phi_\mathrm{in}$}(sub);
  \draw[-] (sub)node[above right]{$\Delta\phi$}-- (dot1);
  \draw[-] (dot1) -- (dot1|-c1_mult) coordinate(temp);
  \draw[-latex] (temp)-- (c1_mult.west);
  \draw[-latex] (dot1)-- (dot1|-c2_mult)--(c2_mult.west);
  \draw[-latex] (c1_mult) -- (sumz1);
  \draw[-latex] (sumz1)--(delay1);
  \draw[-] (delay1) --(dot2);
  \draw[-latex] (dot2) -- (dot2-|sum1) --(sum1);

  \draw[-] (dot2) -- ++(0,0.6)coordinate(temp2) --(temp2 -| sumz1);
  \draw[-latex] (temp2 -| sumz1) -- (sumz1);

  \draw[-] (sum1) -- (dot4) -- (dot4|-sum2);
  \draw[-latex](dot4|-sum2)--(sum2);

  \draw[-] (c2_mult)--(c2_mult-|sum1);
  \draw[-latex] (c2_mult-|sum1) -- (sum1);

  \draw[-latex] (sum2)--(delay2);
  \draw[-] (delay2)-- (dot5);
  \draw[-latex] (dot5) -- ++ (0,.6) coordinate(temp3) -- (temp3-|sum2)--(sum2);
  
  \draw[-latex] (dot5) --(dot5-|sub) --  (sub)node[below left]{$\phi_\mathrm{out}$};

  \draw[-latex] (dot4)--(mult_out);
  \draw[-latex] (mult_out)-- node[above]{$\hat{f}$} (output);

  \node [draw,dashed,inner ysep=2mm, inner xsep=1mm, fit=(dot1)(c2_mult)(temp2)(sum1), color=KITblue, cap=butt] (sc) {};
  \node [draw, dashed,inner ysep=2mm,inner xsep=2mm, fit=(sum2)(delay2)(temp3), color=KITblue, cap=butt] (phase_integrator) {};

  \node [above right, color=KITblue] at (sc.north west) {Loop filter};
  \node [below right, color=KITblue] at (phase_integrator.south west) {Phase integrator};

    \end{tikzpicture}
  
    \caption{Block diagram of the DPLL used for IF estimation}
    \label{fig:block_diagram_pll}
\end{figure}

 A DPLL of second order is depicted in Fig. \ref{fig:block_diagram_pll}. The z-domain transfer function can be easily derived as
\begin{equation}
    H(z)=\frac{\hat{F}(z)}{\Phi_i(z)}=\frac{C_1(z-1)+C_2(z-1)^2}{(z-1)^2+C_2(z-1)+C_1}\mathrm{,}
\end{equation}
where $\Phi_i(z)$  and $\hat{F}(z)$   are the z-domain representation of the phase $\varphi_i$ of the input signal and the estimated frequency $\hat{f}(t)$ respectively. $C_1$ and $C_2$ are the filter coefficients of the IIR loop filter, whose values can be chosen in order to match a certain characteristic frequency $\omega_0$ and a damping factor $\zeta$ according to \cite{shayan}
\vspace{-0.2cm}
\begin{align}
    C_2=2\zeta\omega_0/f_\mathrm{s},\\
    C_1=C_2^2/4\zeta^2,    
\end{align}
under the assumption that $f_\mathrm{s}\gg\omega_0/2\pi$. The overall behavior corresponds to a lowpass filter of second order. Thus, $C_1$ and $C_2$ can be chosen so that the DPLL filters all noise outside of the effectively used Nyquist band for each given information bitrate. The damping factor is set to ${\zeta=1/\sqrt{2}}$ to achieve the flattest transfer function.
\subsubsection{Linear least squares (LLS) instantaneous frequency estimation}
IF estimation by a linear least squares approximation consists of fitting a polynomial of degree $\lambda$ to the unwrapped phase vector  $ \bm{\phi}$ of the received signal in the time domain within a window  of given length $L$ according to 
\begin{equation}
    \bm{\phi}=\bm{T}\bm{a},
\end{equation}
where $\bm{a}$ is the vector of polynomial coefficients and the matrix 
\begin{equation}
   \bm{T}=
    \begin{pmatrix}
        t_0^0 && t_0^1 && \ldots &&t_0^\lambda \\[.3em]
        t_1^0 && t_1^1 && \ldots &&t_1^\lambda\\[.3em]
        \vdots && \vdots && \ddots && \vdots\\[.3em]
        t_L^0 && t_L^1 && \ldots &&t_L^\lambda
    \end{pmatrix}%
\end{equation}
contains the sampling times of the measurement. The least squares solution for the estimated coefficients $\bm{\hat{a}}$ is given by 
\begin{equation}
    \bm{\hat{a}}=\left(\bm{T}^\mathsf{T} \bm{T}\right)^{-1}\bm{T}^\mathsf{T}\bm{\phi}  
\end{equation}
and the derivative with respect to time yields a polynomial for the IF given by 
\begin{equation}
\hat{f}_p[n]=\frac{1}{2\pi}\left(\hat{a}_1+2\hat{a}_2t_n+\ldots+p\hat{a}_\lambda t_n^{\lambda-1}\right)\text{.}
\end{equation}

The length of the moving window is chosen to match the duration of a coded bit ($L_\mathrm{Man}=\frac{M}{2}$, and $L_\mathrm{6b8b}=\frac{3M}{4}$ for the Manchester and 6b8b code respectively), since this is the period in which the IF varies at a constant linear slope. This corresponds to a quadratic behavior of the phase over time (${\lambda=2}$). Nevertheless, a polynomial approximation of a higher degree is necessary to estimate the IF over windows that are not perfectly synchronized with the coded bits. A good empirical compromise between estimation accuracy and avoiding overfitting is a fifth-order polynomial approximation for the phase over time, i.e., $\lambda=5$.

\subsection{Synchronization}
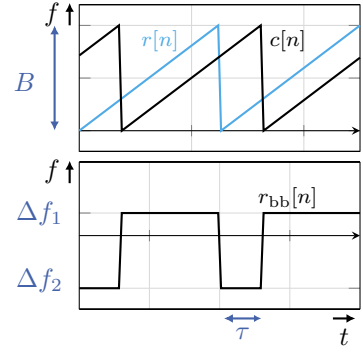
\begin{figure}[tb]
    \centering
        \begin{tabular} [width=1\columnwidth]{c}
        \begin{tikzpicture}[baseline]
        \begin{axis}[
            width=0.6\columnwidth,
            height=0.4\columnwidth,
            axis x line = middle,
            axis background/.style={fill=white},
              grid=major, %
              grid style={gray!30}, %
              y label style={at={(-0.06,0.5)}, font=\normalsize},
              enlarge x limits=false,
              enlarge y limits=false,
              xmin = 0,
              ymax = 0.6,
              ymin = -.1,
              xticklabels={},
              ytick={0,0.25,.5},
              yticklabels={},
              legend style={at={(1,1)}, font=\large, fill opacity=0, text opacity=1, column sep=0.2cm},
              legend cell align={left},
              tick label style={font=\large},
              execute at begin axis={
                            \draw [thick] (rel axis cs:0,0) -- (rel axis cs:1,0)
                                  (rel axis cs:0,1) -- (rel axis cs:1,1);
                        }
            ]
            \addplot+[mark=none, color=KITcyanblue, thick]
        table[x=t,y=x1,col sep=comma] {fmcw_skizze.csv};

        \addplot+[mark=none, color=black, thick]
        table[x=t,y=x2,col sep=comma] {fmcw_skizze.csv};
        
        \end{axis}
   
        \end{tikzpicture}
    \\
  
\begin{tikzpicture}[baseline]
    \begin{axis}[
        width=0.6\columnwidth,
        height=0.4\columnwidth,
        axis background/.style={fill=white},
        axis x line = middle,
          grid=major, %
          grid style={gray!29}, %
          y label style={at={(-0.06,0.5)}, font=\normalsize},
          x label style={at={(0.5,-0.1)}, font=\large},
          xmin = 0,
          xtick={0,0.125,...,1},
          ytick={-0.3571,0.15306},
          xticklabels={},
          yticklabels={},
          ymax = 0.5,
          ymin = -0.5,
          legend style={at={(1,1)},anchor=north east, font=\small,  fill opacity=0, text opacity=1},
          legend cell align={left},
          tick label style={font=\large},
          execute at begin axis={
                            \draw [thick](rel axis cs:0,0) -- (rel axis cs:1,0)
                                  (rel axis cs:0,1) -- (rel axis cs:1,1);
                        }
        ]

        \addplot+[mark=none, color=black, thick]
        table[x=t,y=diff,col sep=comma] {fmcw_skizze.csv};
    \end{axis}

    \end{tikzpicture}

\end{tabular}
\begin{tikzpicture}[remember picture, overlay]

  \draw[KITblue,<->,>=latex, arrows = {Latex[width = 1.2mm, length=1.2mm]-Latex[width = 1.2mm, length=1.2mm]}, thick] (-4.4,0.5) to (-4.4,1.9);
  \draw (-4.8,1.15) node[ color=KITblue]{$B$};
   \draw (-4.6,-0.6) node[ color=KITblue]{$\Delta f_1$};
\draw (-4.6,-1.5) node[ color=KITblue]{$\Delta f_2$};
\draw[KITblue,,<->,arrows = {Latex[width = 1mm, length=1mm]-Latex[width = 1.2mm, length=1.2mm]}, thick] (-1.65,-2) to (-2.15,-2);
\draw (-1.9,-2.2) node[ color=KITblue]{$\tau$};
\draw[->,thick,arrows = {-Latex[width = 1.2mm, length=1.2mm]}] (-0.7,-2) -- node[below]{$t$} (-0.4,-2) ;
\draw[->,thick,arrows = {-Latex[width = 1.2mm, length=1.2mm]}] (-4.2,-0.2) -- node[left]{$f$} (-4.2,0.1) ;  
\draw[->,thick,arrows = {-Latex[width = 1.2mm, length=1.2mm]}] (-4.2,1.9) -- node[left]{$f$} (-4.2,2.2) ;
\draw (-3,1.7) node[ color=KITcyanblue, font=\footnotesize]{$r[n]$};
\draw (-1.3,1.7) node[ color=black, font=\footnotesize]{$c[n]$};
\draw (-1.3,-0.40) node[ color=black, font=\footnotesize]{$r_\mathrm{bb}[n]$};
\end{tikzpicture}
\vspace{0.4cm}
    \caption{Mixing products in the case of unsynchronized signals}
    \label{fig:fmcw_sketch}
\end{figure}

Since the recovery of the information relies on a precise estimation of the IF, the timing synchronization between the unmodulated local oscillator signal $c[n]$ and the received signal $r[n]$ is crucial. A timing mismatch between these signals leads to large and abrupt jumps in the baseband signal. Fig. \ref{fig:fmcw_sketch} sketches the shape of the baseband signal $r_\mathrm{bb}$ for unsynchronized $r[n]$ and $c[n]$. For a known bandwidth and repetition rate of the chirped transmitted signal, some processing steps typically used for range estimation in frequency-modulated continuous wave (FMCW) radar can be used to estimate the timing offset $\hat{\tau}$ between $c[n]$ and $r[n]$.
It can be readily shown that the IF of the mixing products $\Delta f_1$ and $\Delta f_2$ in the downconverted baseband signal $r_\mathrm{bb}[n]$ shown in Fig. \ref{fig:fmcw_sketch} are proportional to the timing offset $\tau$ between the mixed frequency modulated signals. Thus,

\begin{equation} \label{eq:fmcw}
    \hat{\tau}=T_0 \,\frac{\Delta f_1 }{B}=T_0\left(1-\frac{\Delta f_2}{B}\right)  %
\end{equation}
yields an estimate of $\tau$.

Since the total bandwidth $B_0$ and the duration $N$ of a chirp are not altered by the modulation scheme, this synchronization approach does not need any pilot symbols and can be performed using the modulated signal. For an SDR implementation of the proposed system, $N$ samples of the received signal can be buffered for the duration of a chirp and the timing can be corrected according to \eqref{eq:fmcw}.

\section{Theoretical Performance Analysis}\label{sec:CRB}
Since the information is modulated in the time-frequency domain, the effective bit energy and noise variance for decoding do not necessarily correspond to the physical signal-to-noise ratio (SNR) of the received signal. Thus, the influence of the SNR on the detection of symbols and the resulting theoretical optimal performance are derived in this section.

A lower bound of the achievable BER is obtained assuming ideal synchronization and a statistically efficient unbiased estimate of the IF. The Cramér-Rao bound (CRB) of this estimation over an observation of $N_\mathrm{obs}$ samples in the presence of AWGN with variance $\sigma^2$  corresponds to \cite{Boashash}
\begin{equation}
\mathrm{var}[\hat{f}]\geq \frac{12}{\left( 2\pi\right)^2\left(A^2/\sigma^2\right) N_\mathrm{obs} \left(N_\mathrm{obs}^2 -1\right)}\cdot f_\mathrm{s}^2 \mathrm{,}
\end{equation} 
where $A$ represents the constant amplitude of the frequency modulated signal and $f_\mathrm{s}$ is the sampling frequency of the measurement. This variance is the effectively observed noise on the $\hat{f}_\text{bb}$ signal before detection instead of the physical noise power of the AWGN channel. 

For the Manchester code, an observation window of ${N_\mathrm{obs}=M/2}$ corresponds to a fair comparison for the proposed estimators. This window clearly coincides with the length of the LLS estimation. Similarly, the DPLL estimation was optimized to suppress all frequency components higher than the Nyquist band. According to the sampling theorem, this also corresponds to a window of $M/2$ in the time domain. Analogously, a fair comparison for the 6b8b code corresponds to to an observation window of $N_\mathrm{obs}=3M/4$

The energy per bit $E_\mathrm{b}$ is given by the IF deviation of the received signal $r[n]$ from the constant slope chirped signal $c[n]$. For the Manchester code it can be visualized as the area below the $r_\text{bb}[n]$ curve in Fig. \ref{fig:signals} for the duration of a triangular-shaped bit. For a given bit rate, its length and height are $M$ and $B_0M/N$ respectively, with $N$ being the number of samples per chirp repetition. This yields a total bit energy of 

\begin{equation}
    E_\text{b,Man}=\frac{B M^2}{2N}\mathrm{.}
\end{equation}

For the 6b8b code, both the length and height of a modulated bit are a factor $3/2$ larger, which yields a bit energy of $ E_\text{b,6b8b}=\frac{9}{4}\ E_\text{b,Man}$ and leads to an improved performance.

As a result, the bit error probability $P_\mathrm{e,CRB}$ for an estimator that matches the CRB at the output of the matched filter is then  given by \cite{proakis2008}

\begin{equation}
    P_\mathrm{e,CRB}=Q\left(\sqrt{\frac{2E_\mathrm{b}}{\mathrm{var}[\hat{f}]}}\right)\text{,}
\end{equation}
where $Q$ stands for the tail distribution function of the standard normal distribution and $E_\mathrm{b}$ is the bit energy for the used constant-weight code. 
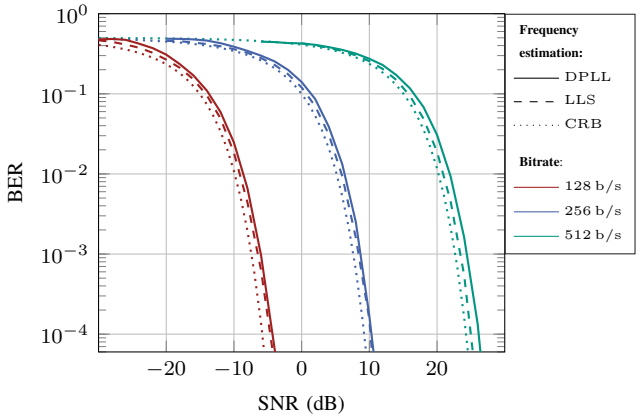
\begin{figure}[tb]
    \centering
    \begin{tikzpicture}

\begin{axis}[%
        width=5.4cm,
        height=4.5cm,
        at={(0.758in,0.645in)},
        scale only axis,
        xtick={-20,-10,0,10,20},
        xmin=-30,
        xmax=30,
        xlabel style={font=\color{white!15!black}, font= \footnotesize},
        xlabel={SNR (dB)},
        ymode=log,
        ymin=0.6e-04,
        ymax=1,
        yminorticks=true,
        ylabel style={font=\color{white!15!black}, font=\footnotesize},
        ylabel={BER},
        axis background/.style={fill=white},
        xmajorgrids,
        ymajorgrids,
        legend pos=south west,
        legend style={at={(1,1)}, anchor=north west, legend cell align=left, align=left, draw=white!15!black, font=\tiny},
        tick label style={font=\footnotesize}
        ]

        \addlegendimage{white} \addlegendentry{
        \hspace{-0.6cm}\textbf{Frequency}}
        \addlegendimage{white} \addlegendentry{
        \hspace{-0.6cm}\textbf{estimation:}}
        \addlegendimage{black} \addlegendentry{$\mathrm{DPLL}$}
        \addlegendimage{black, dashed} \addlegendentry{$\mathrm{LLS}$}
        \addlegendimage{black, dotted} \addlegendentry{$\mathrm{CRB}$}
        \addlegendimage{white} \addlegendentry{}
        \addlegendimage{white} \addlegendentry{\hspace{-0.6cm}\textbf{Bitrate}:}
        \addlegendimage{KITred} \addlegendentry{$\SI{128}{b/\second}$}
        \addlegendimage{KITblue} \addlegendentry{$\SI{256}{b/\second}$}
        \addlegendimage{KITgreen} \addlegendentry{$\SI{512}{b/\second}$}

\addplot[thick, KITgreen,dotted]
        table[x=snr,y=512,col sep=comma] {ber_man_cr_test.csv};      
\addplot[thick, KITgreen,dashed]
        table[x=snr512_lls_man,y=ber512_lls_man,col sep=comma] {postConf_Results.csv};
\addplot[thick, KITgreen, solid]
        table[x=snr512_pll_new,y=ber512_pll_new,col sep=comma] {postConf_Results.csv};

\addplot[thick, KITblue,dotted]
        table[x=snr,y=256,col sep=comma] {ber_man_cr_test.csv};
\addplot[thick, KITblue,dashed]
        table[x=snr256_lls_man,y=ber256_lls_man, col sep=comma] {postConf_Results.csv};
\addplot[thick, KITblue, solid]
        table[x=snr256_pll_man,y=ber256_pll_man,col sep=comma] {postConf_Results.csv};

\addplot[thick, KITred, solid]
        table[x=snr128_pll_man,y=ber128_pll_man,col sep=comma] {postConf_Results.csv};
        
\addplot[thick, KITred,dashed]
        table[x=snr128_lls_man,y=ber128_lls_man,col sep=comma] {postConf_Results.csv};
\addplot[thick,KITred,dotted]
        table[x=snr,y=128,col sep=comma] {ber_man_cr_test.csv};

\end{axis}
\end{tikzpicture}
    \caption{BER simulations for AWGN channel with the Manchester code}
    \label{fig:BER}
\end{figure}
\section{Results and discussion}\label{sec:results}
\begin{figure}[tb]
    \centering
    \begin{tikzpicture}

\begin{axis}[%
        width=5.4cm,
        height=4.5cm,
        at={(0.758in,0.645in)},
        scale only axis,
        xmin=-30,
        xmax=30,
        xlabel style={font=\color{white!15!black}, font=\footnotesize},
        xlabel={SNR (dB)},
        xtick={-20,-10,0,10,20},
        ymode=log,
        ymin=0.6e-04,
        ymax=1,
        yminorticks=true,
        ylabel style={font=\color{white!15!black}, font=\footnotesize},
        ylabel={BER},
        axis background/.style={fill=white},
        xmajorgrids,
        ymajorgrids,
        legend style={at={(1,1)}, anchor=north west, legend cell align=left, align=left, draw=white!15!black, font=\tiny},
        tick label style={font=\footnotesize}
        ]

\addlegendimage{white} \addlegendentry{
\hspace{-0.6cm}\textbf{Frequency}}
\addlegendimage{white} \addlegendentry{
\hspace{-0.6cm}\textbf{estimation:}}
\addlegendimage{black} \addlegendentry{$\mathrm{DPLL}$}
\addlegendimage{black, dashed} \addlegendentry{$\mathrm{LLS}$}
\addlegendimage{black, dotted} \addlegendentry{$\mathrm{CRB}$}
\addlegendimage{white} \addlegendentry{}
\addlegendimage{white} \addlegendentry{\hspace{-0.6cm}\textbf{Bitrate}:}
\addlegendimage{KITred} \addlegendentry{$\SI{128}{b/\second}$}
\addlegendimage{KITblue} \addlegendentry{$\SI{256}{b/\second}$}
\addlegendimage{KITgreen} \addlegendentry{$\SI{512}{b/\second}$}

\addplot[thick, KITred,solid]%
        table[x=SNR_128_pll,y=BER_128_pll,col sep=comma] {BER6b8b.csv};%
\addplot[thick, KITblue,solid]%
        table[x=SNR_256_pll,y=BER_256_pll,col sep=comma] {BER6b8b.csv};%
\addplot[thick, KITgreen,solid]
        table[x=SNR_512_pll,y=BER_512_pll,col sep=comma] {BER6b8b.csv}; %

\addplot[thick, KITgreen,dashed]
        table[x=SNR_512_lls,y=BER_512_lls,col sep=comma] {BER6b8b.csv}; %
\addplot[thick, KITblue,dashed]
        table[x=SNR_256_lls,y=BER_256_lls,col sep=comma] {BER6b8b.csv}; %

\addplot[thick, KITred,dashed]
        table[x=SNR_128_lls,y=BER_128_lls,col sep=comma] {BER6b8b.csv}; %

        \addplot[thick,KITred,dotted]
        table[x=snr,y=128,col sep=comma] {ber_man_cr_6b8b.csv};       
\addplot[thick, KITblue,dotted]
        table[x=snr,y=256,col sep=comma] {ber_man_cr_6b8b.csv};
\addplot[thick, KITgreen,dotted]
        table[x=snr,y=512,col sep=comma] {ber_man_cr_6b8b.csv};   

\end{axis}

\end{tikzpicture}
    \caption{BER simulations for AWGN channel with the 6b8b code}
    \label{fig:BER6b8b}
\end{figure}
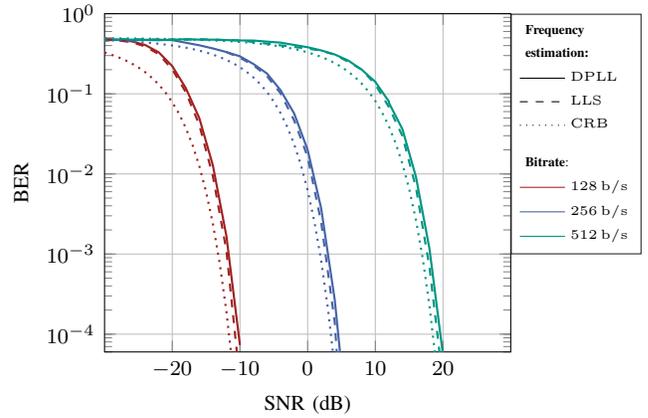

In order to evaluate the performance of the modulation scheme and the complete communication system in a realistic scenario, we carried out Monte-Carlo simulations at different bit rates. The simulations were performed at a sampling rate of $f_\text{s}=2^{16}$ \si{\hertz} since it suits a real implementation on modern software-defined radio devices such as the \textit{\mbox{ETTUS-USRP-E100}}. The chirp bandwidth and repetition rate were set to $B_0=\SI{1024}{\hertz}$ and $R=\SI{4}{\hertz}$, respectively, in order to match the standard specifications. The results for the Manchester code and 6b8b code with each of the studied frequency estimation algorithms are shown in Fig. \ref{fig:BER} and Fig. \ref{fig:BER6b8b} respectively. Additionally, the theoretically optimal bit error probability assuming an efficient unbiased estimator is plotted as a reference.

It can be observed that the theoretically predicted performance is nearly approached when using both DPLL and LLS with the Manchester code. LLS estimation only leads to a marginal improvement of the overall BER performance. This behavior can be explained since it operates at the same effective observation window as the DPLL allowing only marginal gains. As expected, the energy per bit $E_\mathrm{b}$  is increased by a factor $9/4$ with the 6b8b code, which leads to a major improvement of the BER for all bitrates. This performance gain is achieved at the cost of a significantly higher computational effort for the maximum likelihood detection at the receiver.

\section{Conclusion}

In this paper, we propose a novel modulation scheme for efficient communication within the homing signal according to the specifications of modern search and rescue standards. Every building block for a complete SDR implementation of the transceiver is described in detail and its performance is studied numerically and analytically showing near to optimal performance in realistic scenarios. Although the proposed scheme is designed with a very specific context and its specifications in mind, it can be easily adapted to suit integrated communication and sensing using the frequency-modulated waveform in an FMCW radar scenario. This kind of application will be studied in the future.

\section*{Acknowledgement}
This work received funding from the German Federal Ministry for Economic Affairs and Climate Action (BMWK) under grant agreement 20Q1964B.

\bibliographystyle{IEEEtran}
\bibliography{references_asilomar}

\end{document}